\documentclass{JAC2003}
\usepackage{graphicx}
\begin{document}
\title{ELECTRODYNAMICS OF SUPERCONDUCTORS EXPOSED TO HIGH FREQUENCY
       FIELDS}

\author{Ernst Helmut Brandt, Max-Planck-Institut f\"ur
        Metallforschung, Stuttgart, Germany\thanks{ehb @ mf.mpg.de}}

\maketitle

\begin{abstract}
The electric losses in a bulk or film superconductor exposed to a
parallel radio-frequency magnetic field may have three origins:
In homogeneous vortex-free superconductors losses proportional to
the frequency squared originate from the oscillating
normal-conducting component of the charge carriers which is always
present at temperatures $T>0$.
With increasing field amplitude the induced supercurrents approach
the depairing current at which superconductivity breaks down.
And finally, if magnetic vortices can penetrate the superconductor
they typically cause large losses since they move driven
by the AC supercurrent.
\end{abstract}

\section{INTRODUCTION}

The phenomenon of superconductivity was discovered in 1911 by Heike
Kamerlingh-Onnes in Leiden. After he had achieved to liquify helium
at the temperature of $T=4.2 K$ he observed that the resistivity of Hg
became unmeasurably small below some ``critical temperature''
$T_c = 4.15$ K. A sensitive method to measure the residual resistivity
in this ``superconducting state'' is to observe the temporal decay
of the persistent ``supercurrents'' in a ring, say of Pb
($T_c = 7.2$ K), Sn ($T_c = 3.72$ K), or Nb ($T_c = 7.2$ K) by
monitoring the magnetic field generated by the circulating current.
It turned  out \cite{1} that the supercurrent does not decay
measurably, even after several years. Ideally loss-free
superconducting wires may thus be used to build coils which keep
their magnetic field for years, after the windings have been loaded
with current and then are cut short by a superconducting switch.

  Thus, DC currents in a superconductor can flow practically
loss-free if they are not too large.  However, it turned out that
alternating currents (AC) in superconductors are not completely
loss-free, in particular at high frequencies (RF = radio frequencies,
MW = microwave frequencies). There are essentially three effects
which cause energy dissipation during current flow in
superconductors:

(a) Even in ideally homogeneous bulk superconductors an electric
field $E \propto \omega$ (with frequency $\omega/2 \pi$) is
required to accelerate the ``superconducting electrons'',
the Cooper pairs of the microscopic BCS theory \cite{2}. This
electric field also moves the ``normalconducting'' electrons that
are always present at finite temperatures $T > 0$. The dissipated
power of this effect is $\propto E^2 \propto \omega^2$.

(b) When the current density inside the superconductor reaches
the depairing current density $j_{dp}$, the superconducting
order parameter is suppressed to zero at the place where
$j=j_{dp}$. This means superconductivity disappears and electric
losses appear. This nucleation of the normal state typically
occurs at the specimen surface or in the center of Abrikosov
vortices. In particular, when an increasing magnetic field $H_a$
is applied along a superconducting half space $x>0$ one
initially has $j(x) = (H_a/\lambda) \exp( -x/\lambda ) $ when
$j \ll j_{dp}$, but when $H_a$ reaches the thermodynamic
critical field $H_c = \Phi_0 /( \sqrt8 \pi \lambda\xi \mu_0)$
one has $j \approx j_{dp}$ near the surface, thus one has
$j_{dp} \approx H_c/\lambda$. Here
$\Phi_0 = h/2e = 2.07\cdot 10^{-15}$ Tm$^2$ is the quantum of
magnetic flux, $\lambda$ is the magnetic penetration depth,
and $\xi$ is the superconducting coherence length. Within
Ginzburg-Landau (GL) theory, valid near $T = T_c$, the lengths
$\lambda(T) = \kappa \xi(T) \propto (T_c-T)^{-1/2}$
diverge as $T \to T_c$, but the GL parameter $\kappa =
\lambda/\xi$ is independent of temperature $T$.

(c) Large dissipation may be caused by vortices inside the
superconductor. These move under the action of the induced
AC current, which exerts a Lorentz force on the vortices
and causes them to oscillate and dissipate energy. At high
frequencies the amplitude of this oscillation is smaller
then the range of possible pinning forces caused by
material inhomogeneities, e.g., precipitates or defects
in the crystal lattice. This vortex dissipation then
cannot be suppressed by introducing pins.

 For a flat bulk type-II superconductor (defined by
$\kappa > 1/\sqrt2$) in thermodynamic equilibrium it is
favorable that part of the magnetic flux penetrates in form
of Abrikosov vortices when the applied field $H_a$ equals or
exceeds the lower critical field
$H_{c1} \approx (\Phi_0/4\pi \lambda^2 \mu_0)(\ln\kappa +0.5)$
(for $\kappa > 1.5$) and has not yet reached the upper critical
field $H_{c2} \approx \Phi_0/(2\pi \xi^2 \mu_0)$ where
superconductivity vanishes. The penetration of vortices at an
ideally flat surface may be delayed by a surface barrier leading
to a higher penetration field $H_p \approx H_c \ge H_{c1}$
(overheating). On the other hand,
with superconductors of finite size, demagnetization effects
may allow the vortices to penetrate already at much lower fields.
In particular, for a large film of width $w$ and thickness
$d \ll w$, the penetration field is strongly reduced,
$H_p / H_{c1} \approx d/w \dots \sqrt{d/w} \ll 1$ depending
on the edge profile, see below. An infinitely large thin film
will thus be penetrated by any perpendicular magnetic field
component, even if very small.

\section{AC RESPONSE OF VORTEX-FREE SUPERCONDUCTORS}

  The puzzling fact that superconductors may carry loss-free
DC current but AC currents exhibit electric losses,
was explained by the two-fluid model of Gorter and Casimir in
1934 \cite{1}. Later, the microscopic BCS Theory \cite{2}
essentially confirmed the two-fluid picture, giving for its
phenomenological parameters a microscopic interpretation and
explicit expressions.

\subsection{Two-Fluid Model}

   The phenomenological two-fluid model assumes that the total
electron density is composed of the density of superconducting
electrons $n_s$ and that of normal electrons $n_n$, which have
different relaxation times $\tau_s$ and $\tau_n$. Historically,
Gorter and Casimir assumed $n_n \propto t^4$ ($t=T/T_c$) and
$n_s \propto 1-t^4$.  As in the
Drude model \cite{1}, the drift velocity ${\bf v}$ of each of
these two fluids should obey a Newton law,
  \begin{eqnarray}  % 1
   m\, d{\bf v} / dt = e {\bf E} - m {\bf v} / \tau
  \end{eqnarray}
with $m$ and $e$ the mass and charge of the electron.
The total current density ${\bf J} ={\bf J}_s + {\bf J}_n$
is the sum of the supercurrent ${\bf J}_s =e n_s {\bf v}_s$ and
the normal current ${\bf J}_n =e n_n {\bf v}_n$.
(In all other Sections of this paper the current density is
denoted by ${\bf j}$.)  If one assumes $\tau_s = \infty$ one
obtains from Eq.~(1) the first London equation
  \begin{eqnarray}  % 2
  d{\bf J}_s / dt =(n_s e^2/m){\bf E} ={\bf E}/(\mu_0\lambda^2)
  \end{eqnarray}
with $\lambda = (m / n_s e^2 \mu_0)^{1/2}$ the London depth.
In the London gauge where ${\bf E} = - d {\bf A} / dt$
(induction law) Eq.~(2) may be written in form of the second
London equation ${\bf J}_s = (\mu_0 \lambda^2)^{-1} {\bf A}$.
For the normal electrons one may assume $\tau_n \ll 1/\omega$
with periodic electric field $E \propto \exp(i\omega t)$.
This gives for the normal current
 \begin{eqnarray}  % 3
   {\bf J}_n = (n_ne^2\tau_n/m) {\bf E}.
 \end{eqnarray}
Defining the complex conductivity
$\sigma(\omega) = \sigma_1(\omega) -i\sigma_2(\omega)$ by
$J = \sigma(\omega) E \propto \exp(i\omega t)$, one obtains
 \begin{eqnarray} \nonumber % 4
    \sigma_1(\omega) &=& (\pi n_s e^2/m\omega)\, \delta(\omega)
                      + n_n e^2 \tau_n/m ,   \\
    \sigma_2(\omega) &=& n_s e^2/m \omega
    = (\mu_0 \lambda^2 \omega)^{-1} ~\gg \sigma_1 .
 \end{eqnarray}
For a normal conductor this yields $\sigma_1 = \sigma_n$,
$\sigma_2 = 0$, and the skin depth
$\delta_{\rm skin} =(2/\mu_0 \sigma_n \omega)^{1/2}$.
For superconductors at $\omega=0$ the $\delta$-function in
$\sigma_1$ reflects the ideal DC conductivity,
while at finite frequencies the inductive part dominates,
$\sigma_2 \gg \sigma_1$. However, the small dissipative part
$\sigma_1 = n_n e^2 \tau_n/m$ is important since it causes
the AC losses. In the situation with an AC magnetic field
parallel to the superconductor surface, the current is
forced (a current bias as opposed to a voltage bias)
and one has a dissipation per unit volume $\rho J^2 =
{\rm Re}\{1/\sigma\} J^2 \approx (\sigma_1 / \sigma_2^2) J^2
\approx \sigma_1 E^2$ since $\sigma_1 \ll \sigma_2$. The
dissipation is thus proportional to $n_n \omega^2$.

 The sum $\sigma = \sigma_1 -i\sigma_2$
is analogous to a circuit of a resistive channel
$1/R \propto \sigma_1$ in parallel to an inductive channel
of admittance $1/i\omega L \propto \sigma_2$. Below a
frequency $\omega_0 =R/L$ this circuit is mainly inductive
and above mainly resistive. The ratio of the currents in
the two channels is $J_s / J_n = n_s/(n_n \omega \tau)$. This
defines a crossover frequency $\omega \approx (n_s/n_n)
(1/\tau_n) \approx (n_s/n_n) \cdot 10^{11}$ Hz \cite{1}.

  When the superconductor forms the inner wall of a microwave
cavity with incident parallel magnetic field of amplitude
$H_{\rm inc}$, this wave is almost ideally reflected
by the wall since a surface screening current
$J_s = 2H_{\rm inc}$ is induced. The small dissipated power
per unit area is then $P_s = J_s^2 R_s$, where
 \begin{eqnarray}  % 5
 R_s =\delta^{-1} {\rm Re}\{1/\sigma\} =\delta^{-1} \sigma_1
 /|\sigma|^2 \approx \delta^{-1} \sigma_1/\sigma_2^2
 \end{eqnarray}
is the surface resistance (e.g.\ in units $\Omega$). Here
 $\delta = [\,2/\mu_0 (|\sigma| +\sigma_2) \omega\,]^{1/2}$
is the general skin depth reducing to the superconducting
penetration depth $\lambda$ or to the skin depth
$\delta_{\rm skin}$ in the super or normal conducting
limits. For the superconducting wall one has
$R_s \approx \sigma_1 \mu_0^2 \lambda^3 \omega^2/2$ and the
absorbed versus incident power of this wall is \cite{1}
 \begin{eqnarray}  % 6
   {P_{\rm abs} \over P_{\rm inc}} = {J_s^2~ R_s \over
      c \mu_0 H_{\rm inc}^2} = {4 R_s \over
            c \mu_0 }  \approx {1\over Q} .
 \end{eqnarray}
The quality factor $Q$ of the superconducting cavity is
thus inversely proportional to
$R_s \propto Q^{-1} \propto n_n \omega^2$.

\subsection{Microscopic Theory}

  After the BCS theory \cite{2} had given the microscopic
explanation of superconductivity, the complex AC conductivity
was calculated within this weak-coupling theory \cite{3,4}.
In the extreme local limit ($\lambda \ll \xi_0 = \hbar v_F
 /\pi\Delta $ with $v_F$ = Fermi velocity and $\Delta$
 = energy gap; this assumption actually is not satisfied for
type-II superconductors with GL parameter $\kappa > 0.7$), in
the impure limit (electron mean free path $l=v_F\tau \ll \xi_0$),
and for frequencies below the energy-gap frequency
($h\nu < 2\Delta$,
$\nu = \omega/2\pi$) the resulting AC conductivity may be
expressed as two integrals over an energy variable,
 \begin{eqnarray}  % 7
    {\sigma_{1,2} \over \sigma_n} = \int f_{1,2}(\epsilon,
    \Delta, T, \omega)\, d\epsilon \,,
 \end{eqnarray}
where $\sigma_n = ne^2\tau / m = ne^2 l/p_F$ ($p_F=mv_F$ =
Fermi momentum, $n$ = electron density) is the Drude
conductivity in the normal state and $f_1$ and $f_2$ are
some functions. Evaluating these integrals for the case
$\omega \ll T \ll \Delta$ (in units $\hbar = k_B = 1$) one
obtains \cite{5}
 \begin{eqnarray}  \nonumber % 8
  {\sigma_1 \over \sigma_n} &=& {2\Delta \over T} \exp{
  \Big(\! - {\Delta \over T}\Big) } \ln{9T \over 4\omega}\,, \\
  {\sigma_2 \over \sigma_n} &=& {\pi\Delta \over \omega} \,.
 \end{eqnarray}
The dissipative part $\sigma_1$ and inductive part $\sigma_2$
may be written in the form of the two-fluid model:
 \begin{eqnarray}  \nonumber % 9
  \sigma_1 &\approx& n_{qp}\, e^2 l/p_F \,, \\
  \sigma_2 &\approx& n_s\ e^2/m\omega  \,,
 \end{eqnarray}
where $n_{qp}$ is the quasiparticle density (replacing the
normal electron density $n_n$ of the two-fluid model) and
$n_s$ the superconducting electron density,
 \begin{eqnarray}  \nonumber % 10
    n_{qp} &=& n\, {\Delta \over T}\, \exp\Big(\!-{\Delta
         \over T}\Big)\, 2 \ln{9T \over 4\omega}  \,, \\
  n_s    &=& n\ l\, /\, \xi_0  \,.
 \end{eqnarray}
The quality factor $Q$ of the resonator is now
 \begin{eqnarray}  % 11
  Q^{-1} \propto R_s \approx {\textstyle{1\over 2}}
         \mu_0^2 \lambda^3 \sigma_1
         \omega^2 \propto~ n_{qp}\,\omega^2 \,.
 \end{eqnarray}
Since the quasiparticle density
$n_{qp} \sim \exp(-\Delta/T)$ strongly decreases at low
temperatures $T$, $Q$ should increase drastically.
Note that with increasing purity (increasing $l$)
$\sigma_1$ increases but the penetration depth
$\lambda \approx \lambda_{\rm pure} \sqrt{1 +\xi_0/l}$
decreases in Eq.~(11). Thus, maximum $Q$ is reached at some
intermediate, not too high purity of the superconductor.

\subsection{High-Purity Niobium}

    For the high-purity Nb used in the TESLA cavities,
the frequency dependent surface resistance has been computed
by Kurt Scharnberg within the Eliashberg model that extends
the BCS model to strong coupling superconductors \cite{6}.
Strong (electron--phonon) coupling effects change the amplitude
and the temperature dependence of the gap parameter, they
lead to a renormalization (enhancement) of the quasiparticle
mass, which in turn affects the London penetration depth,
and they result in temperature and energy dependent
quasiparticle lifetimes.
The electron-phonon interaction enters in form of the
Eliashberg function $\alpha^2 F(\omega)$ which was taken
from tunneling experiments. A Coulomb pseudopotential
$\mu^* = 0.17$ and a Coulomb cut-off $\omega_c=240$ meV were
used. At sufficiently low $T$ and low $\omega$ of the
incident radiation, inelastic scattering is negligible
and only disorder induced elastic scattering is
important, which is parameterized by the
normal state scattering rate $\Gamma_N = 1/2\tau$. This
is fit to the surface resistance $R_s$ measured at
$\nu = 1.3$ GHz, yielding $\Gamma_N \approx 1$ meV and
$\tau \approx 3\cdot 10^{-13}$ sec. Nonlocal effects
(wave vector $q >0$) were disregarded, which is partly
corrected for by using the lifetime $\tau$ fitted at 1.3 GHz.

   With these assumptions the surface resistance
$R_s \approx {1\over2}\sigma_1 \mu_0^2 \lambda^3 \omega^2$
of high-purity Nb was computed at $T=2$~K.  Note that $R_s$
is related to the reflectivity $r$ of the metal by
$R_s = (Z_0/4)(1-r)$ where $Z_0 = (\epsilon_0 c)^{-1} =$
377 $\Omega$ is the impedance of the vacuum. Starting from
$R_s \approx 20$ n$\Omega$ at 1.3 GHz the resistance rises
to a few $\mu\Omega$ at 600 GHz and then exhibits a large
step at 750 GHz to a value of 15 m$\Omega$. Above this
energy-gap frequency $\nu = 2\Delta /h$ one has
nearly constant $R_s$ till at least 2000 GHz.

\begin{figure}[htb]
\centering
\includegraphics*[width=78mm]{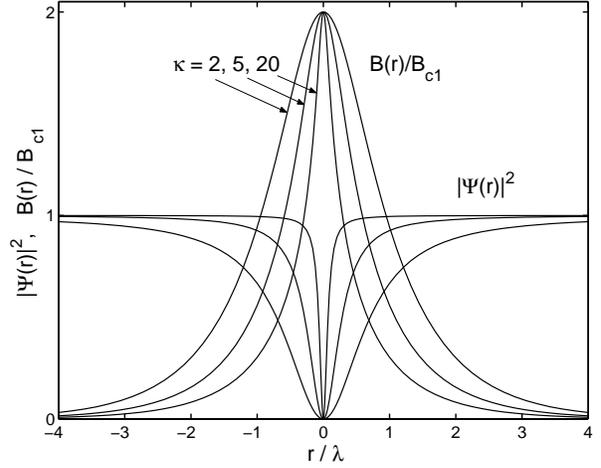}
\caption{Magnetic field $B(r)$ and order parameter $|\psi(r)|^2$
of an isolated flux line calculated from Ginzburg-Landau theory
for GL parameters $\kappa$ = 2, 5, and 20. Note that
the field in the vortex center is $B(0) \approx 2 B_{c1}$.}
\label{f1}
\end{figure}

\begin{figure}[htb]
\centering
\includegraphics*[width=78mm]{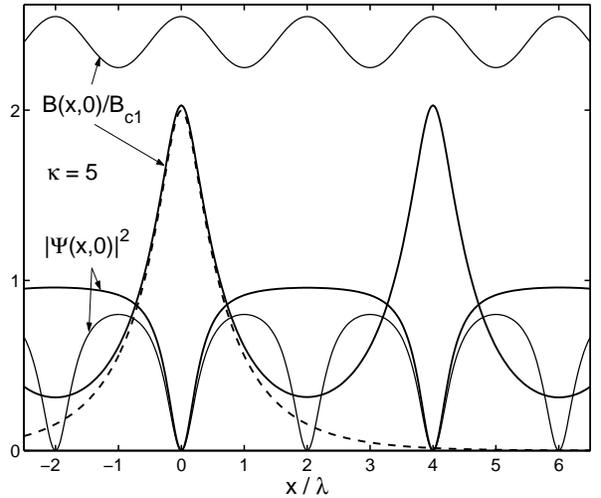}
\caption{ Two profiles of the magnetic field $B(x,y)$ and
order parameter $|\Psi(x,y)|^2$ taken along the $x$ axis
(a nearest neighbor direction) for vortex lattices with
lattice spacings $a=4\lambda$ (bold lines) and $a=2\lambda$
(thin lines). The dashed line shows the magnetic field of
an isolated flux line from Fig.~1. Calculated from
GL theory for $\kappa=5$ \cite{10}.}
\label{f2}
\end{figure}

\begin{figure}[htb]
\centering
\includegraphics*[width=65mm]{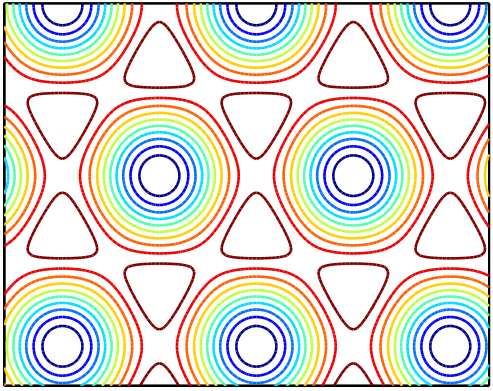}
\caption{Current stream lines, coinciding with the contours
of $B(x,y)$ and $|\psi(x,y)|^2$. Abrikosov solution for the
ideal vortex lattice near the upper critical field $B_{c2}$.}
\label{f3}
\end{figure}

\begin{figure}[htb]
\centering
\includegraphics*[width=80mm]{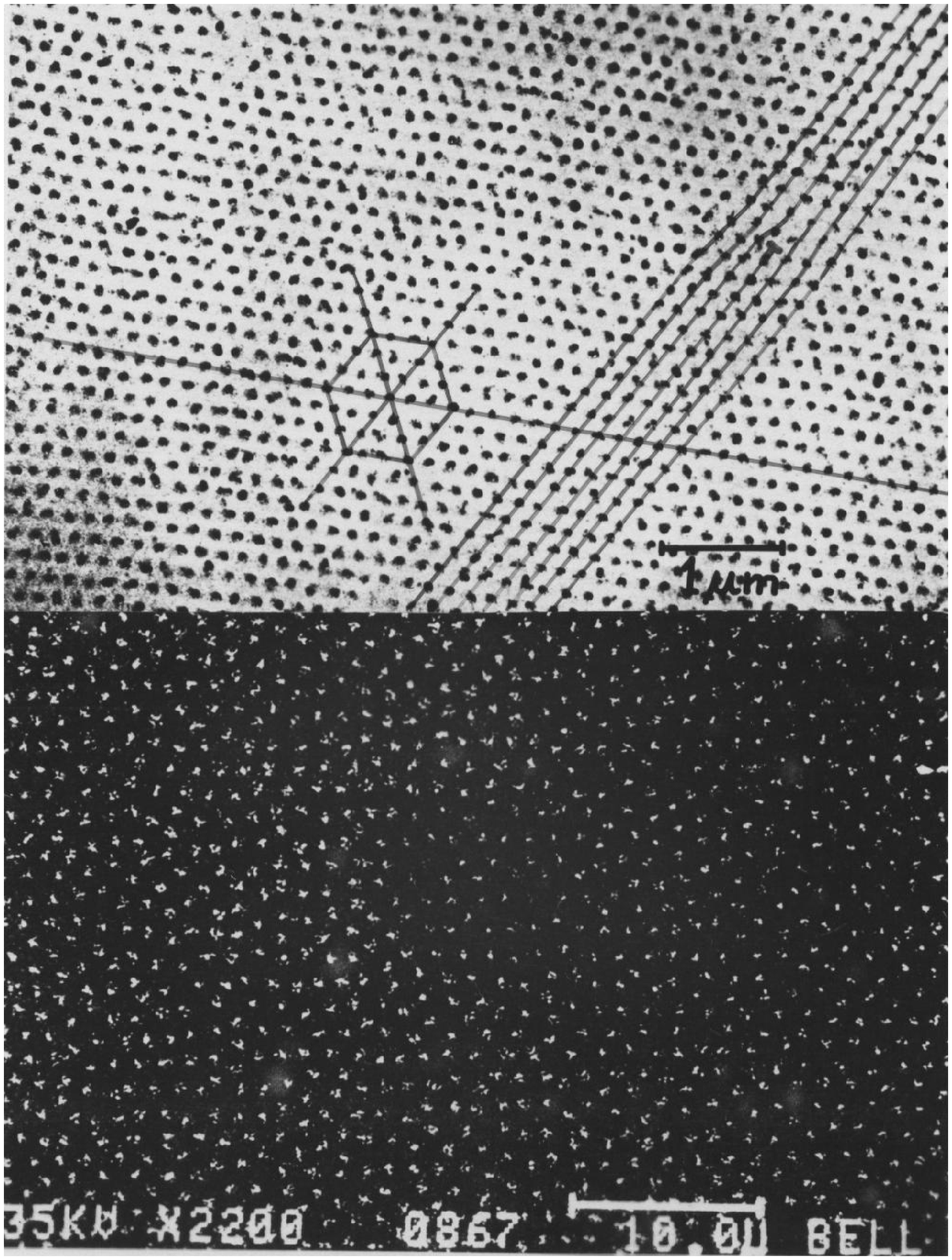}
\caption{Vortex lattice made visible by decoration with
 iron micro-crystallites. Top: Nb disk, 1 mm thick, 4 mm
 diameter, $T=4$ K, $B_a=$ 985 Gauss, vortex spacing
 $a=$ 170 nm  (U.~Essmann and H.~Tr\"auble 1968).
 Bottom: YBa$_2$Cu$_3$O$_{7-\delta}$, $T=77$ K,  $B_a = 20$
 Gauss, $a=1200$ nm (D.~Bishop and P.~Gammel 1987).}
\label{f4}
\end{figure}

\begin{figure}[htb]
\centering
\includegraphics*[width=78mm]{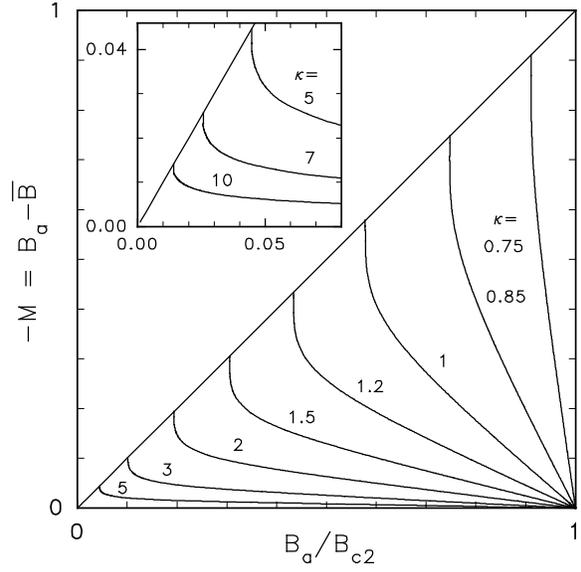}
\caption{Ideal reversible magnetization curves of a long
 superconducting cylinder or slab in parallel field $B_a$
 computed from GL theory \cite{10}.}
\label{f5}
\end{figure}

\begin{figure}[t]
\centering
\includegraphics*[width=40mm]{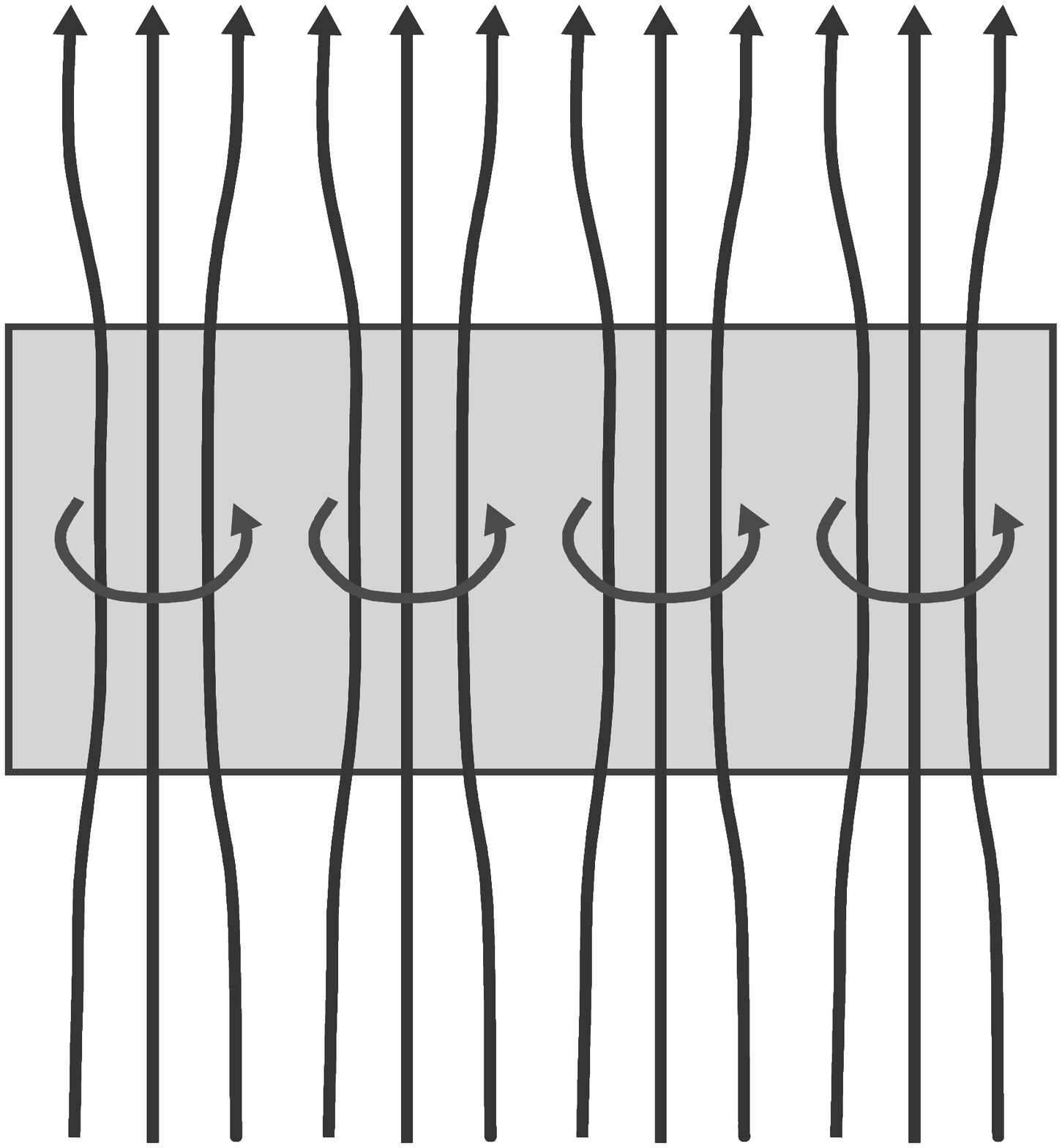}
\caption{Magnetic field lines of vortex lines in and
 near a superconductor of finite size, and the
 circulating super currents (schematic).}
\label{f6}
\end{figure}

\begin{figure}[htb]
\centering
\includegraphics*[width=78mm]{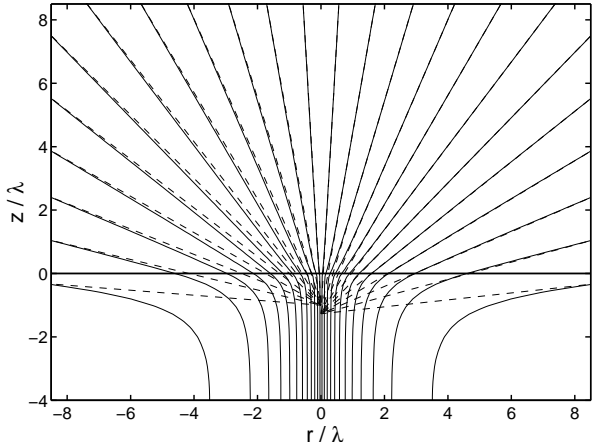}
\caption{Magnetic field lines of a single vortex in a
 superconducting film of thickness $d=8\lambda$ (or half space
 $z \le 0$). Analytical solution of London theory \cite{11}.}
\label{f7}
\end{figure}

\begin{figure}[htb]
\centering
\includegraphics*[width=80mm]{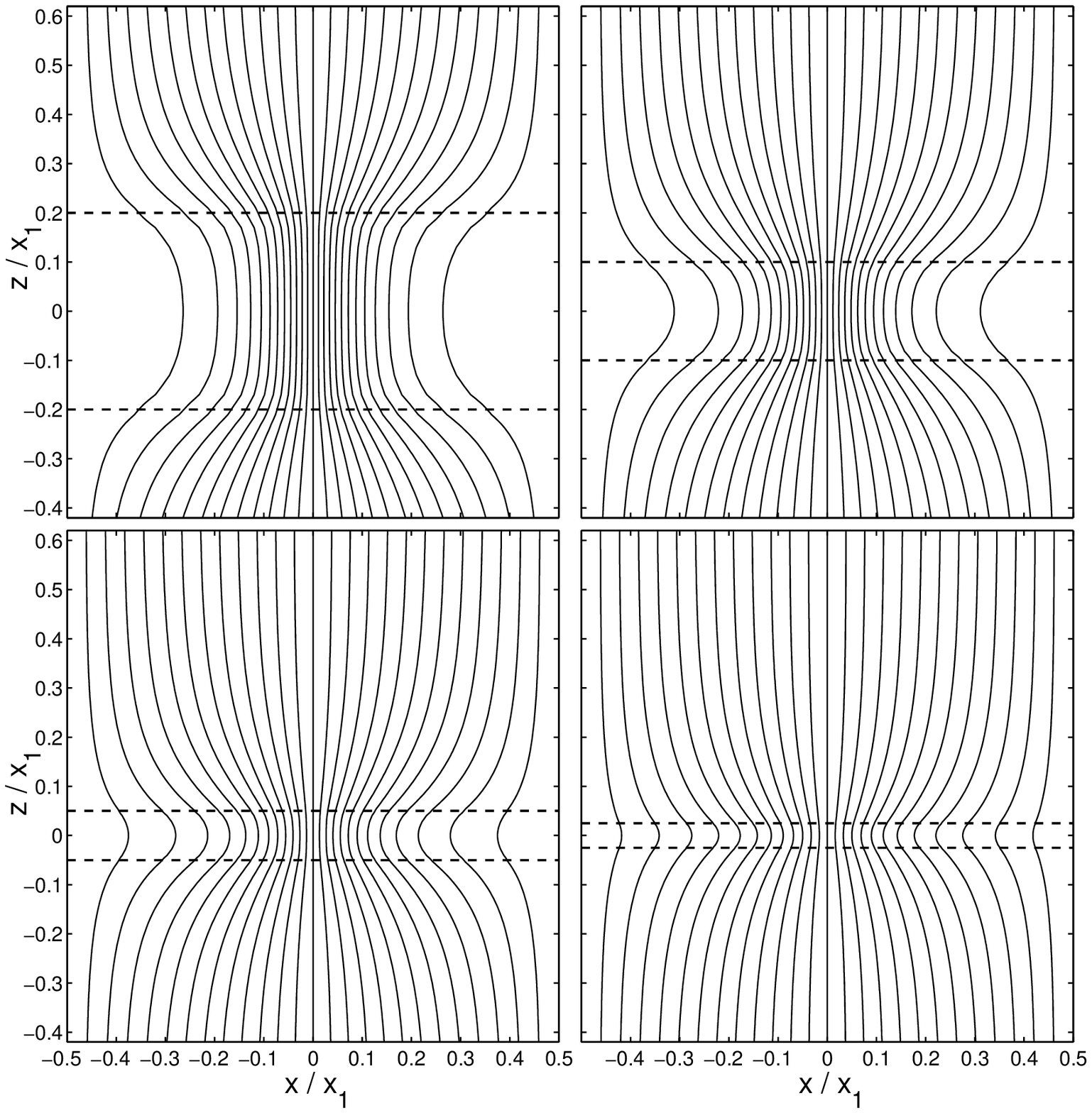}
\caption{Magnetic field lines of the periodic vortex lattice
  in films of thicknesses $d= 4\lambda$, $2\lambda$, $\lambda$,
  and $\lambda/2$. From GL theory for $\kappa = 1.4$ and
  $ \bar B/B_{c2} =0.04$ \cite{12}. The dashed lines mark the
  film surfaces. $x_1$ = vortex spacing.}
\label{f8}
\end{figure}

\section{VORTICES IN SUPERCONDUCTORS}

\subsection{Ginzburg-Landau and London Theories}

  Before the microscopic explanation of superconductivity
was given in 1957 by BCS \cite{2} there were very
powerful phenomenological theories that were able to
describe the thermodynamic and electrodynamic behavior of
superconductors. In 1935 Fritz and Heinz London established
the London theory, see Eq.(2) above, and in 1952 Vitalii
Ginzburg and Lev Landau conceived the Ginzburg-Landau (GL)
theory. The GL theory may be written as a variational
problem that minimizes the spatially averaged GL free
energy density,
  \begin{eqnarray}  % 12
  \big\langle\! -|\psi^2| +{\textstyle{1\over2}} |\psi|^4
   + |(i\xi \nabla\! +\! {\bf A})\psi|^2 +(\lambda
   \nabla\!\times\!{\bf A})^2 \,\big\rangle \nonumber \\
       = {\rm Minimum} \,.
\end{eqnarray}
Here $\psi({\bf r})$ is the complex GL-function, or order
parameter, and ${\bf A(r)}$ the vector potential of the
magnetic induction ${\bf B} = \nabla\times {\bf A}$. The
two lengths are the magnetic penetration depth $\lambda$
(usually taken as unit length) and the GL coherence length
$\xi$; both lengths diverge  as the temperature $T$
approaches the critical temperature $T_c$,
$\lambda \propto \xi \propto (T_c -T)^{-1/2}$. Their ratio,
the GL parameter $\kappa = \lambda/\xi$ within GL theory
(valid near $T_c$) is independent of $T$.

  The GL theory can be derived from the microscopic BCS
theory (L.~P.~Gor'kov 1959) in the limit $T_c-T \ll T_c$,
yielding for the GL function
$\psi({\bf r}) = \Delta({\bf r})/\Delta_{\rm BCS}$
where $\Delta$ is the energy gap function.
The London theory follows from GL theory in the cases
when the magnitude of the order parameter is nearly constant,
$| \psi | \approx 1$. This condition is fulfilled when
$\xi$ is small as compared to the specimen extension and
to $\lambda$, requiring $\kappa \gg 1$. An arrangement of
straight or arbitrarily curved vortex lines positioned at
${\bf r}_\nu(z)=[x_\nu(z), y_\nu(z), z]$
($\nu = 1,2,3 \dots$) then has a magnetic field that obeys
the London equation modified by adding $\delta$ functions
centered at the vortex cores,
  \begin{eqnarray}  % 13
 (1 -\lambda^2 \nabla^2\,)\, {\bf B(r)} = \Phi_0
 \sum_\nu \int\! d{\bf r}_\nu \, \delta_3({\bf r-r}_\nu)\,.
  \end{eqnarray}

\subsection{Ideal Vortex Lattice}

  In 1957 Alexei Abrikosov, a thesis student of Lev Landau
in Moscow, obtained a periodic solution of the Ginzburg-Landau
equations and recognized that this corresponds to a lattice
of vortices of supercurrent, circulating around each zero of
the order parameter and carrying a quantum of magnetic flux
$\Phi_0$; these vortex lines (or flux lines, fluxons) are
energetically favorable when the
applied magnetic field is between a lower and a higher critical
field, $H_{c1} \le  H_{c2}$ (see Introduction and below). This
solution exists in bulk superconductors with GL parameter
$\kappa \ge 1/\sqrt2$, called type-II superconductors. For this
theoretical discovery Abrikosov received the Nobel Prize
in Physics in 2003.

  Figure 1 shows the magnetic field $B(r)$ and the order
parameter $|\psi(r)|^2$ of one isolated vortex line for three
values of the GL parameter $\kappa =$ 2, 5, 20. One can see
that $B$ decays over the length $\lambda$ and the vortex core
has a radius $ \approx \xi$. For such not too small values
of $\kappa$ to a good approximation the vortex field is
the London solution, with the central singularity smoothened
over the core radius $r_c \approx \sqrt 2\, \xi$ \cite{7,8},
%($K_0(x)$ is a modified Bessel function),
  \begin{eqnarray}  % 14
  B_{v}(r)\! &\approx& {\Phi_0 \over 2\pi \lambda^2}~
    K_0\big({ \sqrt{r^2 +r_c^2} \over\lambda} \big) ~~~
                                           \nonumber\\
  K_0(x)\!\! &=& \left\{ \begin{array} {lr} ~\ln(1.123/x),
                            & x\ll 1  \\[.5mm]
    \sqrt{\pi/ 2x}\, \exp(-x), & x\gg 1 \end{array} \right. .
  \end{eqnarray}
$K_0(x)$ is a modified Bessel function.
The interaction energy of two vortices at a distance
$x\gg \xi$ is $U_{\rm int} = \Phi_0 B_v(x)/\mu_0$.

Figure 2 shows cross sections of $B(x,y)$ and $|\psi(x,y)|^2$
along the nearest neighbor direction $y=0$ of the ideal
triangular vortex lattice for two values of the average induction
$\bar B =\langle B \rangle$ (with vortex spacing $a=2\lambda$ and
$a=4\lambda$) and $\kappa = 5$. The dashed line is $B(r)$ for the
isolated vortex. Figure 3 shows the contour lines of $B(x,y)$
near $B_{c2}$ for the triangular vortex lattice. These lines
coincide with the contours of $|\psi(x,y)|^2$ and with the
stream lines of the supercurrents.

  The vortex lattice was first observed in the electron
microscope by U.~Essman and H.~Tr\"auble \cite{9} at our
Max Planck Institute in Stuttgart, by decoration of the
surface of a Nb disk with ``magnetic smoke'' generated by
evaporating an iron wire in a He atmosphere of 1 Torr, see
Fig.~4. The magnetization $-M = B_a - \bar B$ of the
superconductor calculated numerically \cite{10}  as
function of the applied magnetic field $B_a$ is depicted
in Fig.~5 for ideal (pin-free) long
superconductor cylinders or slabs in parallel $B_a$
(i.e., in absence of demagetization effects) with
various $\kappa = 0.707 \dots 10$. At $\kappa = 1/\sqrt2$
one has $B_{c1} = B_c = B_{c2}$ and the  curve
$M(B_a)$ is the same as for type-I superconductors
(with $\kappa < 1/\sqrt2$), namely, $\bar B=0,~-M=B_a$
(no penetrated flux) for $B_a < B_{c2}$ and $\bar B=B_a,~M=0$
(complete penetration of flux) for $B_a > B_{c2}$.

  When the superconductor is not a long cylinder or slab
in parallel field, demagnetization effects shear the
magnetization curves of Fig.~5 and reduce the field of
first vortex penetration, see below. The vortices end at
the upper and lower surface of finite-size specimens
and send their magnetic field lines into the surrounding
vacuum, see Fig.~6. The resulting modulation
of $B(x,y)$ just outside the surface can be observed by
decoration (Fig.~4) and by magneto-optics or Hall probes.
The magnetic field lines of one vortex in a thick film
in a perpendicular magnetic field are depicted in Fig.~7
as obtained from London theory in \cite{11}.
Figure 8 shows the field lines of the periodic vortex
lattice in films of thicknesses $d= 4\lambda$, $2\lambda$,
$\lambda$, and $\lambda/2$ as calculated in \cite{12}.

\begin{figure}[thb]
\centering
 \includegraphics*[width=50mm]{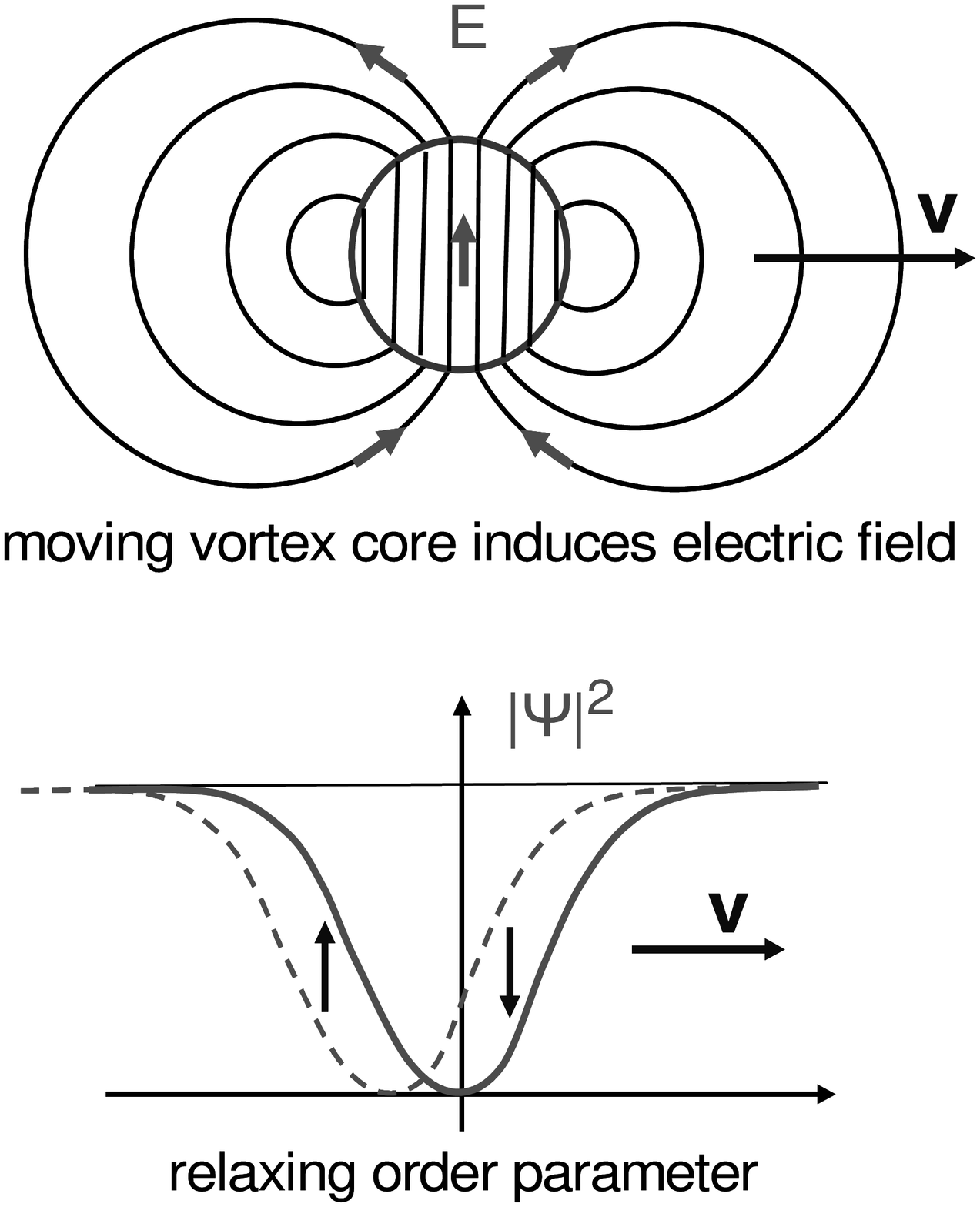}
\caption{Visualization of the origin of energy dissipation
when a vortex moves with velocity $v$. Top: the dipolar
electric field lines induced by this motion run also
through the normal conducting core (Bardeen-Stephen model).
Bottom: During motion of the vortex core the order parameter
relaxes (Tinkham term).}
\label{f9}
\end{figure}

\subsection{Losses by Moving Vortices}

   When a supercurrent flows in a superconductor, either
applied by contacts or caused by a gradient or curvature
of the local magnetic field, this current density ${\bf j}$
exerts a Lorentz force ${\bf f}= {\bf j \times \hat z} \Phi_0$
on a vortex. The Lorentz force density on a vortex lattice
is ${\bf F} = {\bf j \times \bar B}$. Neglecting a small Hall
effect, the vortices move along this force with velocity
${\bf v} = \eta^{-1} {\bf F}$ where $\eta$ is
a drag coefficient or viscosity. The vortex motion induces
an average electric field
 \begin{eqnarray}   % 15, 16
 {\bf E} &=& {\bf\bar B \times v}= \eta^{-1} {\bf\bar B\times
    (j\times\bar B)} = \rho_{\rm ff} {\bf j} \,, \\
 \rho_{\rm ff} \! &\approx&  (\bar B/B_{c2})\ \rho_n.
 \end{eqnarray}
Here $\rho_{\rm ff}$ is
the flux-flow velocity, which at large average inductions
$\bar B$ is comparable to the normal resistivity of the
superconductor at that temperature (measurable by
applying a large field $B_a > B_{c2}$). However, when
only a few vortices have penetrated ($\bar B \ll B_{c2}$)
one has much smaller resistivity $\rho_{\rm ff} \ll \rho_n$.
But even then the vortex-caused dissipation at low $T$ is
typically much larger than the dissipation caused by
the normal excitations.

  Where does this resistive dissipation come from? There are
two effects of comparable size, see Fig. 9. First,
as pointed out by Bardeen and Stephen \cite{1,13}, the motion
of the magnetic field induces a dipolar electric field that
drives current through the superconductor and through the
vortex core. If the vortex core is modelled as a normal
conducting tube of radius $r_c \approx \xi$, the normal
currents inside the vortex core dissipate energy that leads
to the $\rho_{\rm ff}$ of Eq.~(16). Second, as stated by
Tinkham \cite{1,14}, the moving vortex core means that at
a given position the order parameter $|\psi|^2$ goes
down and up again when the core passes. If one assumes
a delay of the recovery of $|\psi|^2$ by a relaxation time
$\tau \approx \hbar / \Delta$ one obtains an additional
dissipation of the order of Eq.~(16). These two sources
of losses are nice for physical understanding. In the
exact calculation of the dissipation of a moving vortex
lattice from time-dependent GL theory \cite{15} these
two sources cannot be separated but the approximate
Eq.~(16) is essentially confirmed \cite{16}, also
by microscopic theory \cite{17}. The numerical and
also the measured flux-flow resistivity in the middle
between the exact values $0$ and $\rho_n$ is somewhat
smaller than the Eq.~(16), i.e., for constant current
source the real dissipation is lower.

\begin{figure}[thb]
\centering
\includegraphics*[width=80mm]{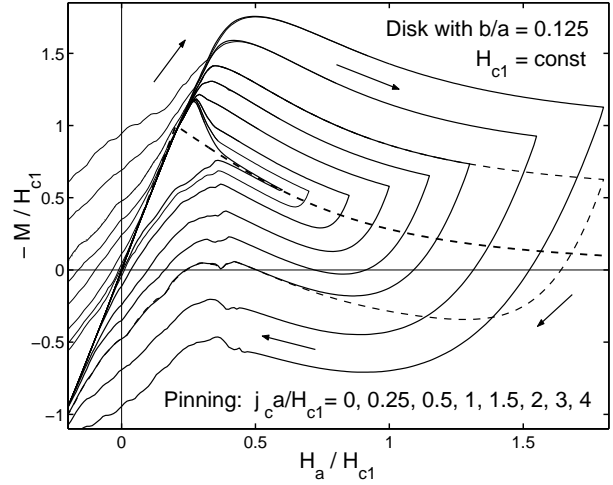}
\caption{Irreversible magnetization curves $M(H_a)$ for a
 disk with radius $a$ and thickness $2b$, $b/a=0.125$,
 for various pinning strengths measured by the parameter
 $a j_c/H_{c1}$. The large hysteresis loop
 belongs to strong pinning $a j_c/H_{c1} =4$. The small
 central loop is for the pin-free disk, whos vortex
 distribution is shown in Fig.~11. The reversible
 magnetization curve of a pin-free ellipsoid with same
 initial slope as $M(H_a)$ is shown as dashed line.}
\label{f10}
\end{figure}

\begin{figure}[thb]
\centering
\includegraphics*[width=70mm]{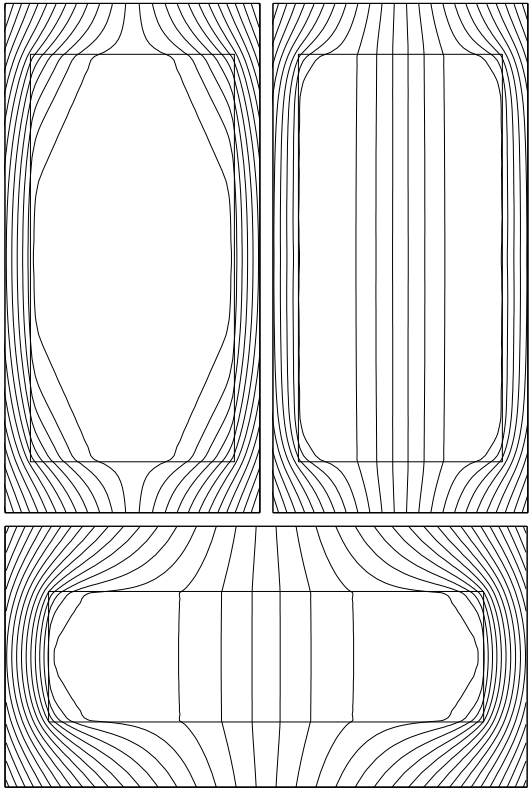}
\caption{Penetration of vortex lines into pin-free cylinders
 with  radius $a$ and height $2b$. Top: $b/a=2$.
 Bottom: $b/a=0.3$. Forced by the applied field $H_a$, the
 vortices enter from the corners, but only when the applied
 field has reached some threshold field do they jump to the
 middle leaving a vortex-free zone near the surface. With
 further increasing $H_a$ the vortices eventually fill the
 cylinder uniformly from the middle. This delayed penetration
 without pinning is called geometrical barrier. Such a
 barrier is absent only for ellipsoid-shaped specimens.}
\label{f11}
\end{figure}

\begin{figure}[thb]
\centering
\includegraphics*[width=80mm]{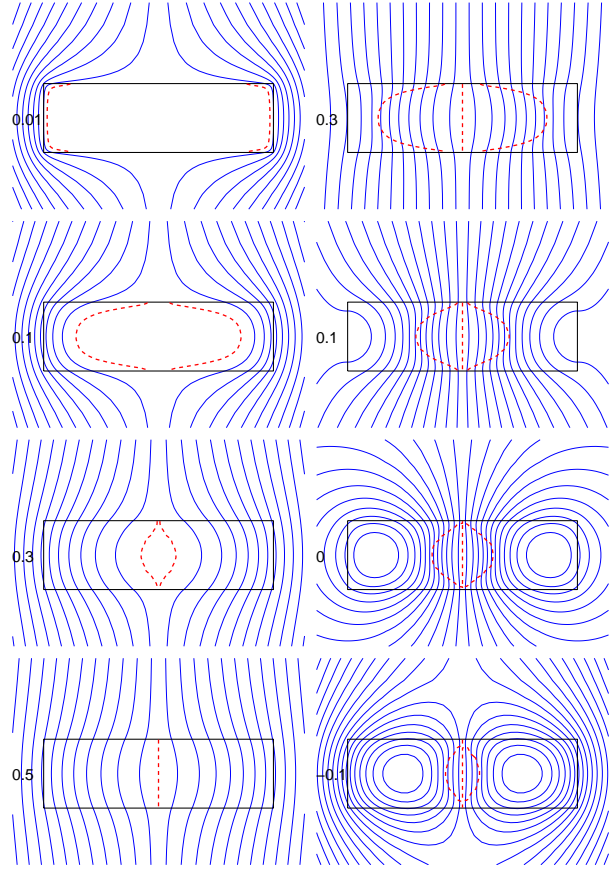}
\caption{Bean model with constant critical current
 density $j_c$ for a superconducting bar with rectangular
 cross  section $2a \times 2b$ ($b/a =0.35$) put into a
 perpendicular  magnetic field $H_a$  that first
 increases from 0 to 0.5 (left column) and then
 decreases again (right column). The parameter
 0.01, 0.1, 0.3, 0.5, 0.3, 0.1, 0, -0.1 is $H_a/(aj_c)$.
 Shown are the magnetic field lines (solid lines) and the
 penetrating fronts (dashed lines) where the current density
 $j$ (flowing along the bar) jumps from $\pm j_c$ to 0 (in the
 field-free and current-free core) or from  $j_c$ to $-j_c$
 (after full penetration of flux).}
\label{f12}
\end{figure}

\begin{figure}[thb]
\centering
\includegraphics*[width=80mm]{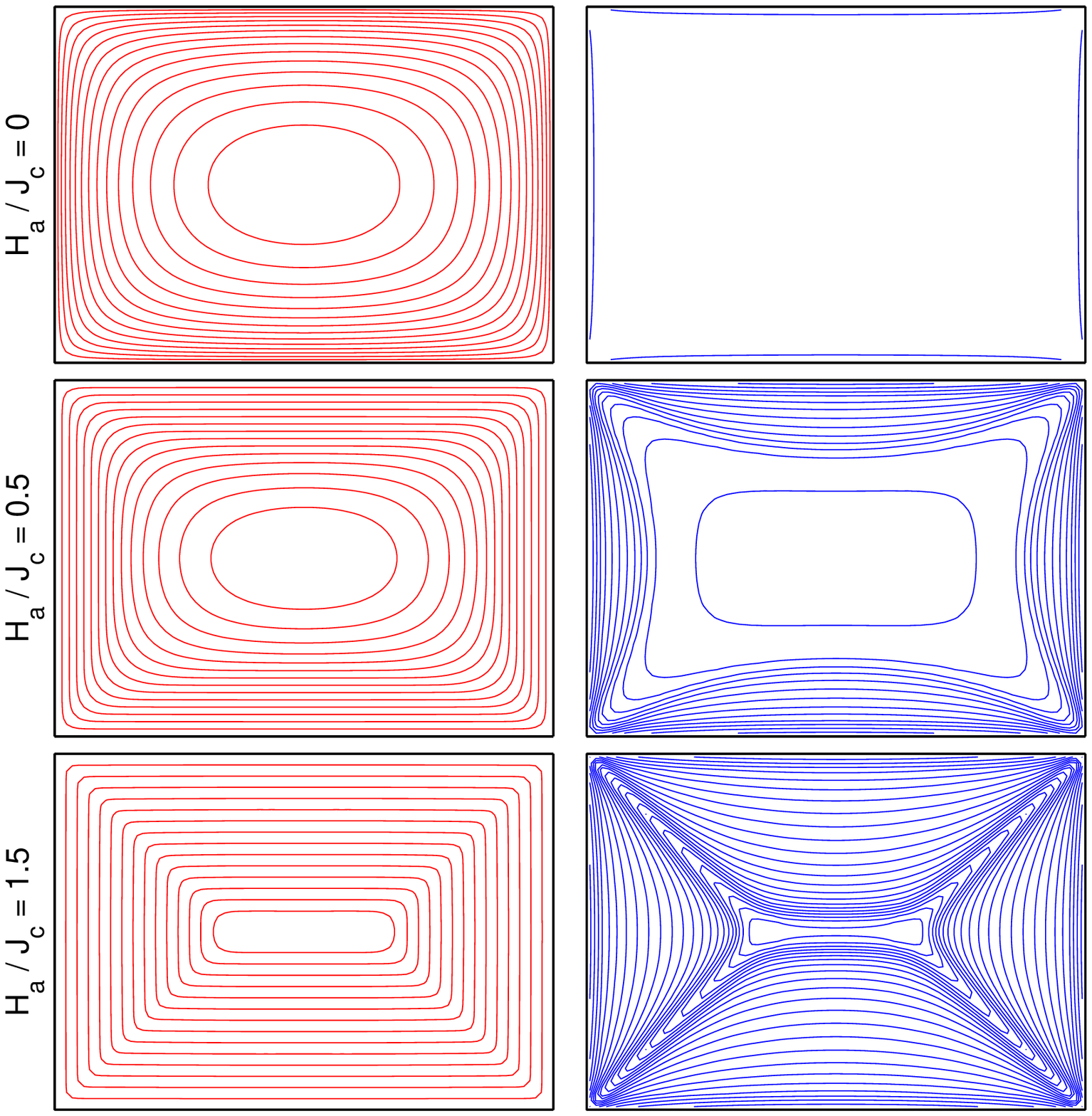}
\caption{Bean model for the penetration of a perpendicular
 magnetic field $B_a=\mu_0 H_a$ into a thin rectangular film
 with thickness $d \ll $ width. The parameter
 $H_a/J_c = 0$, 0.5, 1.5  measures  $H_a$ in units of the
 critical sheet current $J_c=dj_c$. Shown are the stream
 lines of the sheet current in the film,
 ${\bf J}(x,y) = \int {\bf j}(x,y,z)\, dz$ (left), and the
 contour lines of the magnetic field $B_z(x,y)$ in the
 plane $z=0$ of the film (right).}
\label{f13}
\end{figure}

\subsection{Pinning of Vortices}

  When the material is inhomogeneous on the microscopic
length scale of the vortex core $\xi$, then the vortices
are pinned and cannot move as long as the Lorentz force
does not exceed the pinning force, or the current density
$j$ is smaller than the critical current density  $j_c$,
see the reviews \cite{18,19,20}.
In this way the electric losses caused by flux flow can be
avoided, and completely loss-free conductors of DC current
can be tailored by introducing appropriate pinning centers
into the material, e.g., precipitates and crystal lattice
defects. For AC currents small losses remain, however.
One source of AC dissipation is due to the (albeit small)
concentration of normal carriers or excitations and can
be understood from the two-fluid model as discussed above.
The other source is the oscillation of vortices in the
pinning wells. At small displacements $u$ from their
equilibrium position one may assume linear elastic
binding of the vortices to the pins, with a force density
$-ku$. Adding to this the viscose drag force
$-\eta \dot{u}$ and the Lorentz force one obtains the
force balance equation in an AC current
${\bf j}_{ac} \propto \exp(i\omega t)$,
  \begin{eqnarray}   % 17
  {\bf j}_{ac} \times {\bf \bar B} = -k{\bf u}
     -i \omega \eta {\bf u} \,.
  \end{eqnarray}
One can see that at frequencies above $k/\eta$, of
order $\omega/2\pi > 10^7$ Hz, the viscose force
dominates \cite{21}.
Pinning thus cannot prevent vortex oscillations at
high frequencies. This vortex-caused dissipation
increases as $\omega^2$, like the quasiparticle
dissipation, cf.\ Eq.~(11).

   Interesting theoretical problems are the statistical
summation of random pinning forces to obtain the critical
current density $j_c$, and the problem of thermally activated
depinning \cite{18,19,20}. The latter leads to vortex motion
even at small currents densities $j < j_c$ due to finite
temperature. This flux creep may be described by a highly
nonlinear resistivity. In particular, a logarithmic
dispersive activation energy for depinning,
$U(j) = U_0 \ln(j_c/j)$, leads to an often observed
power-law current--voltage curve,
  \begin{eqnarray}   % 18
  E(j) = E_0 \exp[-U(j) / k_BT] = E_0 (j / j_c)^n
  \end{eqnarray}
with a creep exponent $n = U_0/k_B T$. For $n=1$ one has
Ohmic behavior [free flux flow, Eq.~(15)], for
$n\gg 1$ one has flux creep, and in the limit $n\to \infty$
this power law yields the Bean model, in which $j$ is either
$0$ or $j_c$: When at some position one has $j>j_c$, the
vortices rearrange immediately such that $j$ is reduced to
$j_c$ again. This concept is useful for DC currents and at
not too high frequencies where the pinning forces exceed
the viscose drag force.

\subsection{Geometry Effects}

  The electromagnetic properties of a superconductor (and of
any conductor or isolator) depend not only on the material
but also on the geometry of the problem, i.e., on the shape
of the specimen and on the way a magnetic or electric field
is applied. For example, the reversible magnetization curves
of a pin-free superconductor in Fig.~5 apply to the unrealistic
case of very long slabs or cylinders in exactly parallel
field, where demagnetization effects are absent. For the
still unrealistic situation of a perfect ellipsoidal shape
one may calculate from these ideal curves the reversible
magnetization curves of any ellipsoid by using the concept
of the demagnetization factor. But when the specimen shape
is not an ellipsoid, then even for a pin-free superconductor
the magnetization curves have to be computed numerically,
since now the induction (or vortex density) inside the
specimen is no longer spatially constant.

  It turns out that even without pinning such magnetization
curves in general show a hysteresis, i.e., they are
irreversible and depend on the magnetic history, see Fig.~10.
This irreversibility is due to a geometric barrier \cite{22,23}
for the penetration of vortices as illustrated in Fig.~11 for
cylinders (or long bars) with rectangular cross section: When
the applied uniform field $H_a$ is increased, vortex lines
enter at the corners, pulled by the screening currents
(Meissner currents) that flow at the surface, and held back
by their line tension (like a rubber band). With increasing
$H_a$ the vortices penetrate deeper and become longer.
When the vortices from two corners meet at the equator,
they connect and form one long vortex line that contracts
and immediately jumps to the specimen center. During this
rapid jump all their elastic energy is dissipated by the
viscose drag force $F=\eta v$, see text above Eq.~(15).
With further increasing $H_a$ more vortices jump to the
center, crossing the flux-free zone near the surface,
and eventually the entire specimen is filled with vortices
coming from the growing central zone.

   Flux penetration thus occurs with a threshold, over a
``geometrical barrier''. The sudden onset of flux penetration
to the center leads to the sharp maximum in the small (inner,
pin-free) hysteresis loop of $M(H_a)$ in Fig.~10. When
$H_a$ is decreased again, the vortices leave the specimen
essentially without barrier, and at $H_a=0$ all vortices
have left, i.e., one has $\bar B = 0$ and also $M=0$
(since no screening currents flow anymore).
The perpendicular field at which the first vortices enter
at the corners of a pin-free long strip and a circular
disk, both with rectangular cross section $2a \times 2b$,
was computed in \cite{23}:
  \begin{eqnarray}    % 19
 H_{\rm pen}^{\rm strip}&\approx& H_{c1} \tanh\sqrt{0.36\,b/a}\,,
                             \nonumber \\
 H_{\rm pen}^{\rm disk} &\approx& H_{c1} \tanh\sqrt{0.67\,b/a}\,.
   \end{eqnarray}

  In the presence of pinning the hysteresis loops of
$M(H_a)$ in Fig.~10 become larger. The area of such loops
is the energy dissipated during one cycle due to
depinning of vortices. When $H_{c1}$ is negligibly small
as compared to $H_a$, the hysteresis curves and the vortex
density and currents in a superconductor with pinning may be
computed by treating it as a nonlinear conductor, Eq.~(18).
Figure 12 shows how the magnetic field lines (and vortices)
penetrate and exit a thick disk with pinning when an axial
$H_a$ is first increased beyond the field of full penetration,
and then is decreased again \cite{24,25}. The chosen large
creep exponent $n=50$ practically reproduces the Bean model.

   Figures 10, 11, and 12 were computed by time-integration
of an equation for the (scalar) current density $j$ inside
the superconductor; this method implicitly accounts for the
infinitely extended magnetic stray field outside
the specimen, without need to compute it and to cut it off.
From the resulting current density the magnetic field lines
are then easily calculated by the Biot-Savart law.

   A completely different geometry is shown in Fig.~13, namely,
the current stream lines and the contours of the magnetic field
$B_z(x,y)$ in a thin film or platelet of rectangular
shape \cite{26} with pinning and large
creep exponent $n=50$ corresponding to
the Bean model like in Fig.~12. An increasing magnetic field
$H_a$ is applied perpendicular to the film. Initially, when
$H_a \ll J_c = dj_c$ is small, no magnetic flux penetrates
the film, i.e., the circulating screening currents generate a
magnetic field that in the film area is constant (of size $-H_a$)
and exactly compensates the applied field $H_a$. With increasing
$H_a$, magnetic flux penetrates mainly from the middle of the
sides of the rectangle (not from the corners), leaving still
a flux-free zone in the middle. At and beyond some field of
full penetration the current stream lines are concentric
rectangles of constant distance, since the magnitude of the
sheet current has saturated to the constant value $J_c = dj_c$.
The magnetic field has then penetrated to the center, and the
contour lines of $B_z(x,y)$ do not change anymore with
further increasing $H_a$.

\begin{figure}[t]
\centering
\includegraphics*[width=80mm]{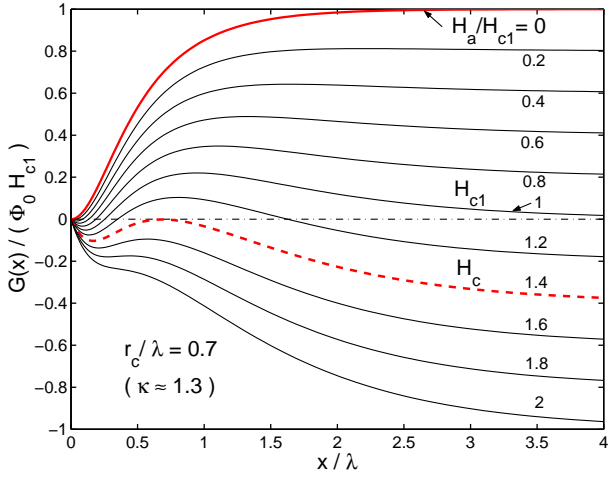}
\caption{Gibbs free energy $G$ of one vortex penetrating into
a superconducting half space to a depth $x$, Eq.~(22).
Parameter is the applied field $H_a$ in units of $H_{c1}$.
The Bean-Livingston Barrier exists for $H_a < H_c$.}
\label{f14}
\end{figure}

\begin{figure}[thb]
\centering
\includegraphics*[width=47mm]{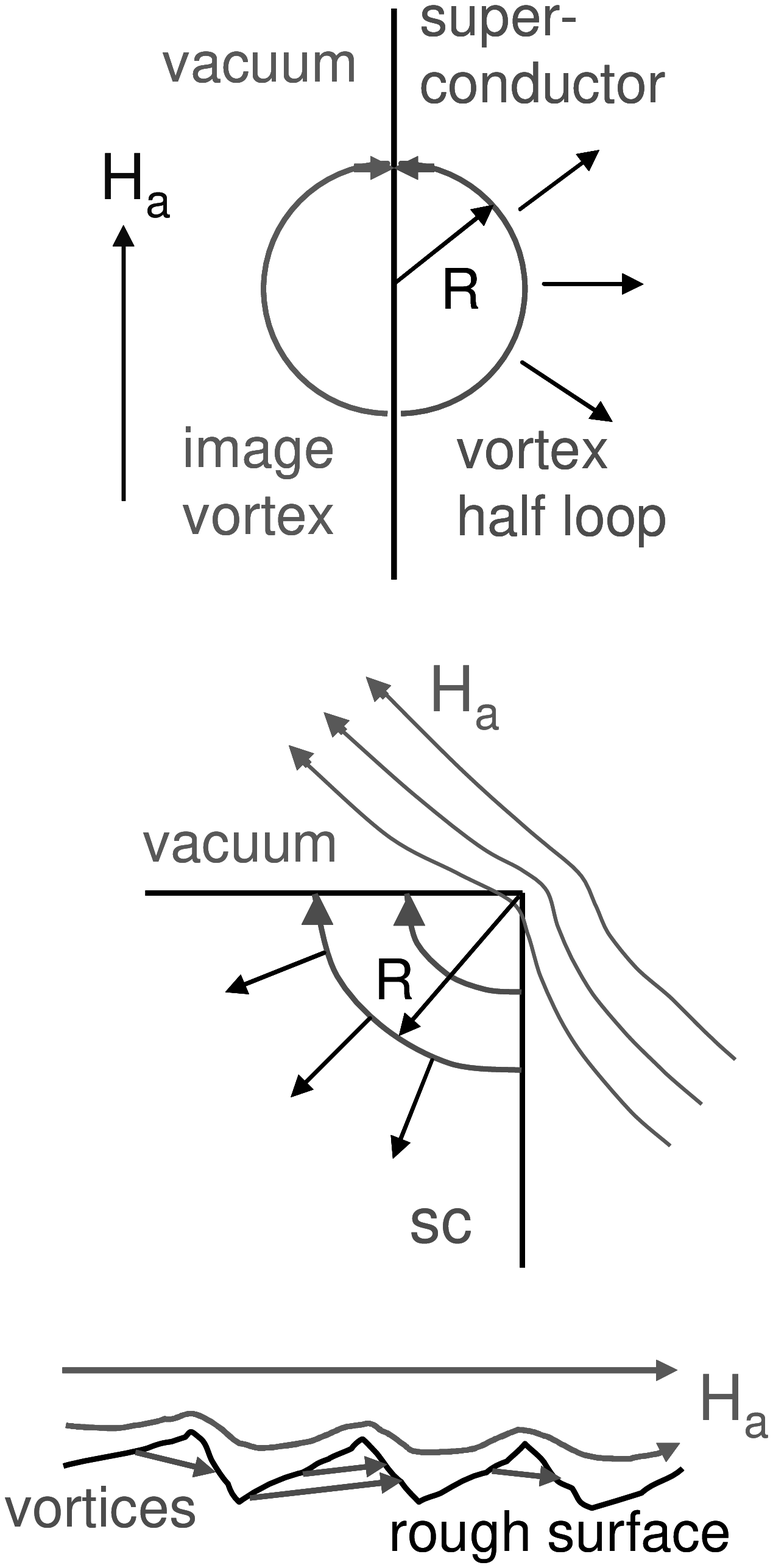}
\caption{Nucleation of vortices as an arc of a circle
 at a planar surface (top), at a rectangular corner (middle),
 and at a rough surface (bottom, schematic).}
\label{f15}
\end{figure}

\begin{figure}[thb]
\centering
\includegraphics*[width=80mm]{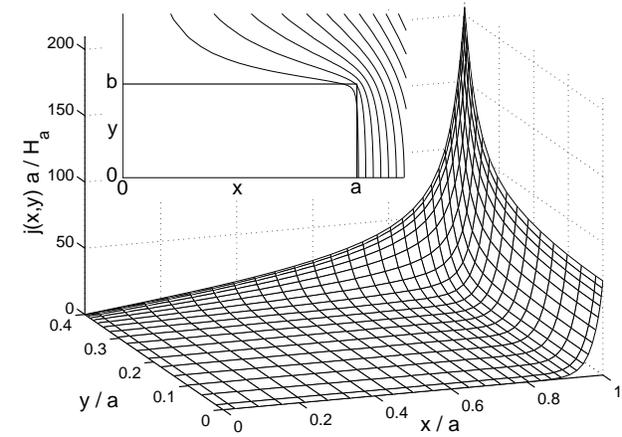}
\caption{Supercurrents $j_z(x,y)$ in a bar with rectangular
 cross section  $2a \times 2b$ ($b/a=0.4$) in the Meissner
 state with  London  penetration depth $\lambda = 0.025 a$.
 The currents (along the bar) are generated by a perpendicular
 applied uniform magnetic field $H_a \| z$ that penetrates to
 a depth $\lambda$. Shown is one quarter of the cross section.
 Note the high (but finite) peak of $j_z(x,y)$ at the corners.
 The inset shows the magnetic field lines.}
\label{f16}
\end{figure}

\begin{figure}[thb]
\centering
\includegraphics*[width=50mm]{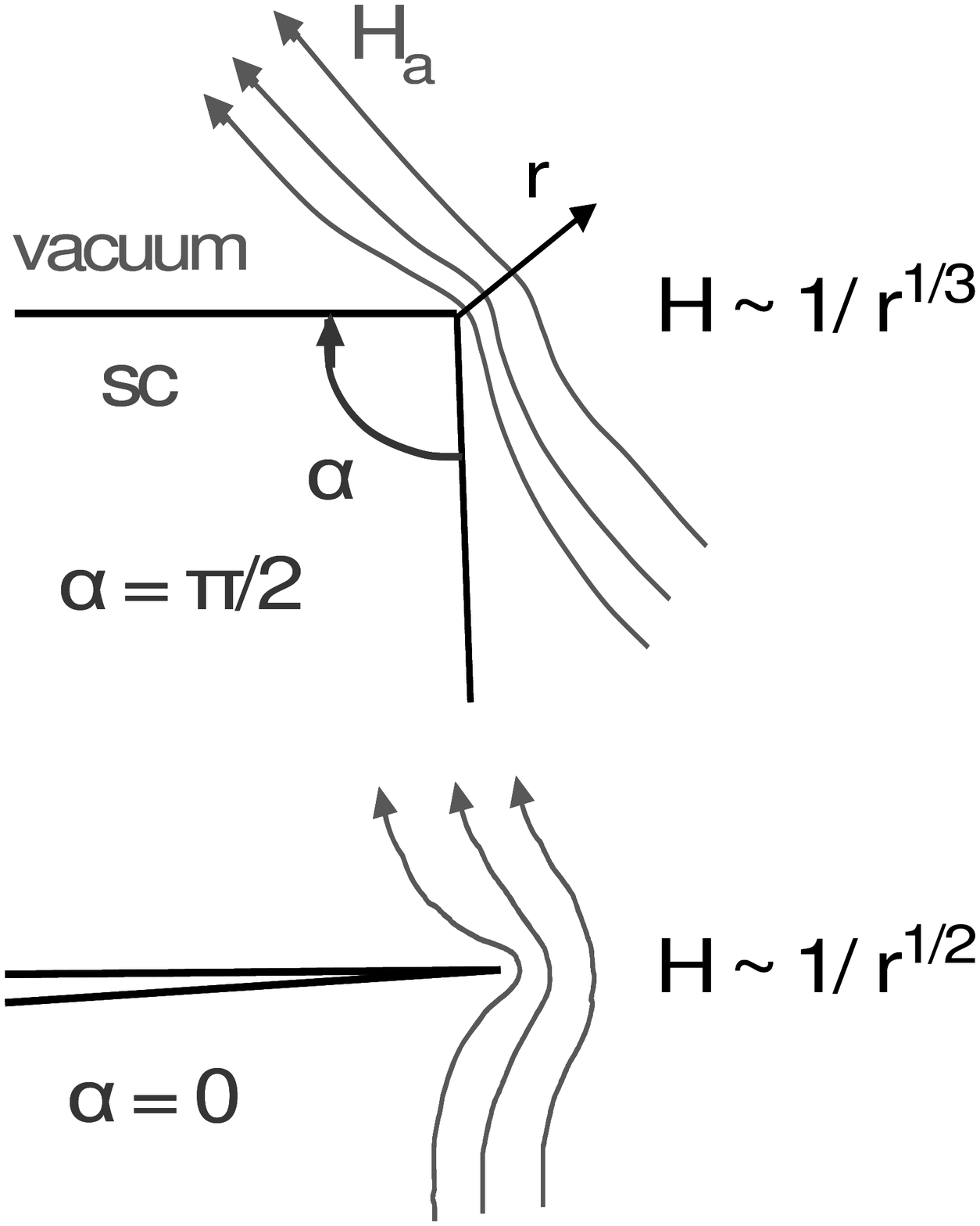}
\caption{Field enhancement near the sharp edge of an
 ideal diamagnet.}
\label{f17}
\end{figure}

\begin{figure}[thb]
\centering
\includegraphics*[width=68mm]{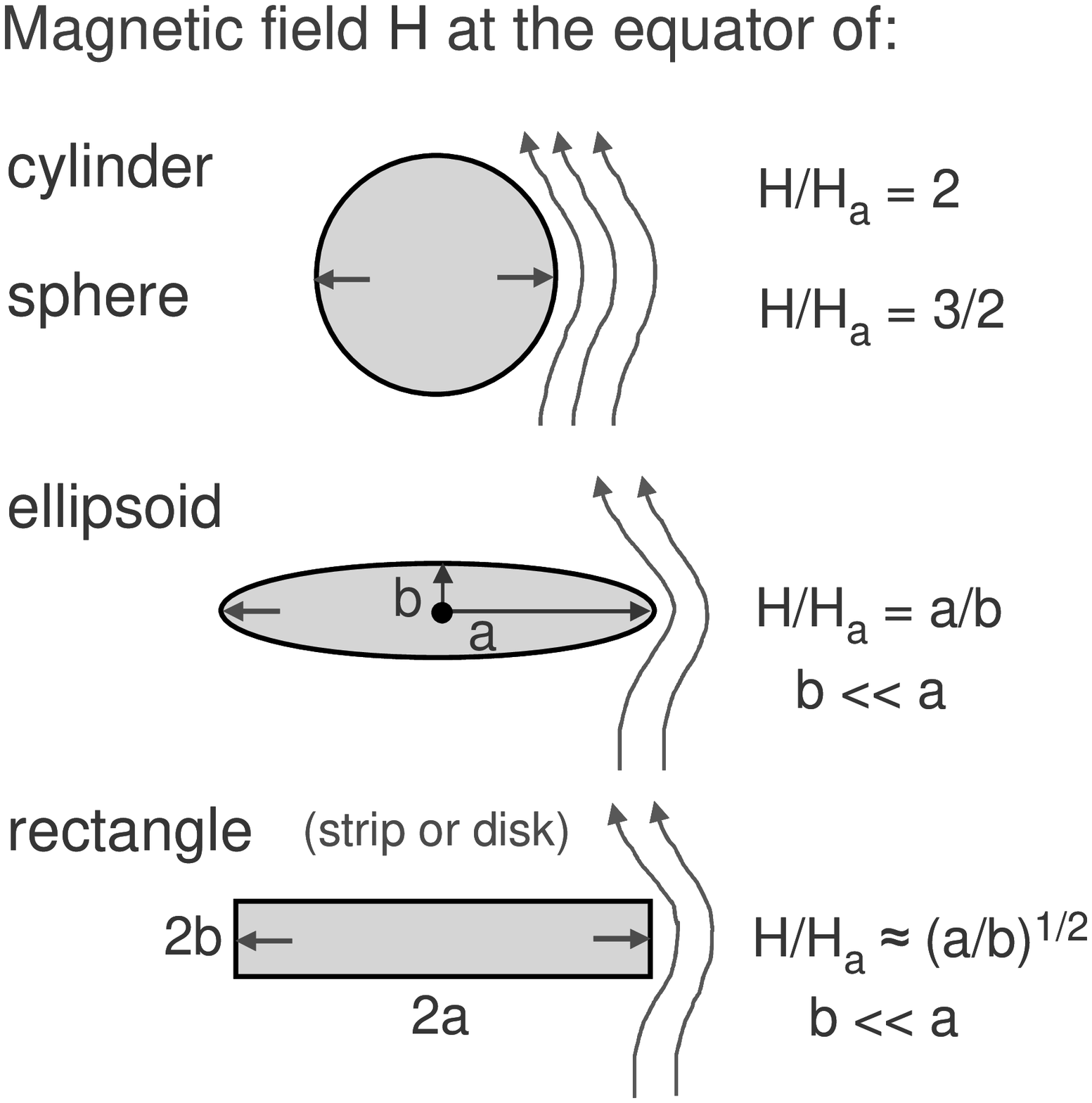}
\caption{Enhancement of the magnetic field $H$ at the equator
 of ideal diamagnetic cylinders, spheres, and strips of
 elliptic or rectangular cross section put in a uniform
 axial applied magnetic field $H_a$.}
\label{f18}
\end{figure}

\subsection{Penetration of First Vortex}

  An important question for RF superconductivity is under
what circumstances and at which applied magnetic field
$H_p$ the first vortex enters the superconductor, since
the presence of even a few vortices can cause large losses.
First I summarize the expressions for the three critical
fields which for type-II superconductors (with
$\kappa \ge 1/\sqrt2$) obey $B_{c1} \le B_c \le B_{c2}$:
  \begin{eqnarray}   % 20
  B_{c1}\! &\approx&\! {\Phi_0 \over 4\pi\lambda^2}\,
          (\,\ln \kappa +\alpha) \,,      \\ \nonumber
  B_{c}\, &=&\! {\Phi_0 \over \sqrt 8\pi \lambda \xi } =
          {\sqrt 2 \kappa \over \ln\kappa +\alpha}\, B_{c1}
                                  \,,     \\ \nonumber
  B_{c2}\! &=&\! {\Phi_0 \over 2\pi\xi^2}
            ~~~~=~ \sqrt2\kappa\, B_c \,, \\ \nonumber
  \alpha(\kappa)\!\! &=&\! {1\over 2}+{1 +\ln 2\over 2\kappa
  -\sqrt2 +2} = \left\{ \!\! \begin{array} {ll} 1.35, & \!
  \kappa=0.71 \\[.5mm] 0.50, & \!\kappa \gg 1~. \end{array} \right.
  \end{eqnarray}
While the thermodynamic ($B_c$) and upper ($B_{c2}$) critical
fields are exact, the lower critical field $B_{c1} =\mu_0 H_{c1}$
has to be calculated numerically from the self energy of a vortex
of length $L$, $U_{\rm self} = \Phi_0 H_{c1} L$. The function
$\alpha(\kappa)$ is
a good analytical fit to the numerical result of \cite{10}. At
$H_a = H_{c1}$ the nucleation of a vortex and motion to a depth
$x \gg \lambda$ does not cost energy, see Fig.~14. However, the
penetrating vortex has to surmount a barrier such that the
field of first penetration $H_p$ is larger than $H_{c1}$.
This barrier was first predicted by Bean and Livingston
(BL) \cite{27} for a superconductor with planar surface in a
parallel applied field $H_a$. The Gibbs free energy $G(x)$ for
this case reads
 \begin{eqnarray}   % 21
  G(x) = {\Phi_0} \Big( H_a e^{-x/\lambda}-
  {\textstyle{1\over 2}} H_v(2x) + (H_{c1}\! -\! H_a) \Big) .
 \end{eqnarray}
In it the first term is the interaction of the vortex with the
applied field $H_a$ or with its screening currents
$(H_a/\lambda) e^{-x/\lambda}$, the second term is the
interaction with the image vortex (at position $-x$, of opposite
orientation), and the third term is an integration constant.
Using the fact that for not too small $\kappa$ one has
$ B_v(0) \approx 2B_{c1}$ (see Fig.~1) one has with Eq.~(14),
$B_{c1} \approx (\Phi_0 / 4\pi\lambda^2) K_0(r_c/\lambda)$
yielding with $B_{c1}$ (20) a core radius
$r_c \approx \xi \exp[-\alpha(\kappa)]$.
With this we may write $G(x)$ (21) in the dimensionless form
 \begin{eqnarray}   % 22
   { G(x) \over \Phi_0 H_{c1}} \approx  {H_a\over H_{c1}}
   \big(e^{-x/\lambda}\! -\!1\big) +\! 1\! -\!{K_0(\!\sqrt
   {4x^2\! +\!r_c^2} /\lambda) \over K_0(r_c / \lambda) }
  \end{eqnarray}
that is plotted in Fig.~14 for $\kappa\approx 1.3$. Of
course, this $G(x)$ is a only approximate, in particular
at small $\kappa$, for which vortex penetration has to be
computed numerically. Anyway, Fig.~14 shows that vortex
penetration becomes favorable at $H_a=H_{c}$ and that the
Bean-Livingston barrier vanishes at $H_a \approx H_c>H_{c1}$.

  The assumption of BL that the entering vortex is long,
straight, and exactly parallel to a planar surface is not
very realistic. Alternatively, one may assume that the first
vortex nucleates and penetrates in form of a small loop, say
a half circle of radius $R$, see Fig.~15 top. The
self-energy of this half circle is approximately
$U_{\rm self} = \pi R (\Phi_0^2 / 4\pi\lambda^2 \mu_0)$,
putting the outer cut-off radius $\approx R$ instead of
$\Lambda \gg R$ in the logarithm
$\ln(\lambda/\xi) \to \ln(R/\xi) \approx 1$ when $R$
is of order of $\xi$. The interaction
of this vortex loop with the surface screening current of
density $j_s$ is $U_{j_s} \approx (\pi R^2 /2) \Phi_0 j_s$
(flux quantum times loop area times $j_s$). For a planar
surface one has $j_s = H_a/\lambda$ directly at the surface.
The criterion that $U_{j_s} \ge U_{\rm self}$ at
$H_a \ge H_p$ yields then
 \begin{eqnarray}   % 23
   H_p \approx {\Phi_0/\mu_0 \over 2\pi\lambda R} =
    {\sqrt2 \xi \over R} H_c \approx H_c \,,
 \end{eqnarray}
which is just the BL result. Thus, the assumption of a
penetrating vortex loop does not change much the
penetration field of a planar surface.

   However, when the surface has roughness with
characteristic length $\ge \xi$, then vortices will
penetrate at sharp points or cusps, see Fig.~15. At a corner
with angle $\alpha = 90^o$, the screening current directly
at the surface is strongly enhanced at this corner; Fig.~16
shows this for a superconducting bar with square cross section
$2a \times 2a$ and penetration depth $\lambda = 0.025 a$, to
which a uniform transverse $H_a$ is applied. A rough
estimate gives for this geometry an enhancement of the
screening current at this corner, $j_s = C H_a/\lambda$,
by a factor $C\approx 4$. The field of first vortex
penetration $H_p$ is then reduced from Eq.~(23) by just this
factor,  $H_p \approx H_c/C \approx H_c/4$.

  For sharper
corners the enhancement of $j_s$ and reduction of $H_p$
are even larger. As shown in the textbook of
Landau-Lifshitz (Electrodynamics of Continua) for
an ideal diamagnetic material at a corner with angle
$\alpha$ (Fig.~17) the magnetic field diverges as
$H\propto 1/r^\beta$ with exponent
$\beta = (\pi -\alpha) /(2\pi -\alpha)$, where $r$ is
the distance to the point of the corner. This gives
$H \propto 1/r^{1/3}$ for $\alpha = \pi/2$ and
$H \propto 1/r^{1/2}$ for $\alpha \to 0$.

  Similarly, an axially applied magnetic field flowing
around an ideal diamagnetic cylinder, sphere, or disks
with elliptical or rectangular cross section of aspect
ratio $b/a \ll 1$, is enhanced at its equator by
factors 2, 3/2, $a/b$, or $\approx (a/b)^{1/2}$,
respectively, due to the strong curvature of the
field lines at this line, see Fig.~18.

\subsection{Vortices in thin films}

  One has to distinguish two quite different types of
vortices in thin film superconductors: vortices
perpendicular or parallel to the film plane. In wide thin
films with width $w=2a \gg$ thickness $d=2b$, the vortices
will nearly always run perpendicular across the film
thickness, even in tilted applied field $H_a$, because of
the large demagnetization factor of this film. This means
the circulating currents prefer to flow in the film plane.
Only when $H_a$ is exactly parallel to the film surface, or
when the film is coating a bulk superconductor that screens
any perpendicular field component, then vortices parallel
to the film plane may occur.

  When the film is of finite size, one may use Eqs.~(19)
to estimate at which applied perpendicular field component
$H_{az}$ the first vortices penetrate, namely already at a
very small field, smaller than $H_{c1} \sqrt{d/w}$. When
the film edges are wedge-shaped or sharp, the penetration
field is even smaller, cf.\ Fig.17 and Fig.~18 (elliptical
edge). Into infinitely extended or closed films
 (e.g., a Nb layer covering the
inner surface of a Cu cavity) any perpendicular field
will penetrate since the field lines cannot flow around
the film. Only when this film has holes or slits can some
magnetic flux cross the film via these holes, but the magnetic
field in the holes will be larger than $H_{az}$ by at least
the ratio of film area over the total area of all holes.
However, the field in the holes will penetrate into the
film when it is of the order of $H_{c1}$ times the square
root of film thickness over hole distance. Thus, even
such a perforated pin-free film will be penetrated by a
perpendicular field that is very much smaller than $H_{c1}$.
The peaked magnetic field, circulating current, and pair
interaction of perpendicular vortices in thin films were
calculated for infinitely extended \cite{28} and
finite-size (e.g. rectangular) films \cite{26,29,30}.

  Pinning of vortices will not appreciably enhance all
these penetration fields at high radio frequencies, where
the (elastic) pinning forces are smaller than the viscose
drag force. If the small applied perpendicular magnetic
field is a DC field (e.g., some stray field or the
earth magnetic field) then the additional RF field will
even favor the penetration of the DC field in form of
vortices, since it ``shakes'' the vortices. As shown in
\cite{31,32}, shaking of vortices by an AC field oriented
perpendicular to the vortices leads to the relaxation of
irreversible currents if the AC amplitude exceeds some
threshold value. This vortex creep means that even in very
small $H_{az}$, perpendicular vortices will penetrate under
the action of a large-amplitude RF field, and then these
vortices oscillate and dissipate energy.

  The problem of parallel vortex lines in a thin film
with $d \ll \lambda$ was solved by Alexei Abrikosov (1964),
Vadim Shmidt (1969), and in an elegant way by
Alex Gurevich \cite{33}. The lower critical field is
enhanced in thin films as compared to bulk superconductors,
 \begin{eqnarray}   % 24
   B_{c1} = { 2\Phi_0 \over \pi d^2} \Big( \ln{d\over
     \lambda} -0.07\Big) \,,
 \end{eqnarray}
and the field at which the surface barrier for vortex
penetration disappears is also enhanced,
 \begin{eqnarray}   % 23
   B_p =  {\Phi_0 \over 2\pi d \xi}  \,.
 \end{eqnarray}
For example, a NbN film with $\xi=5$ nm, $d=20$ nm has
$B_{c1} = 4.2$ T and $B_p = 6.37$ T, much better than
the penetration field $B_p \approx B_c =0.18$ T for Nb
at low $T$.

To enhance the operating RF amplitude in microwave cavities
for accelerators and reduce the losses, Gurevich \cite{34}
suggests to use solid Nb or Pb with multilayer coating
on its inner surface by alternating superconducting
and insulating layers with $d< \lambda$. This will prevent
penetration of vortices into the bulk superconductor when
the vortex penetration field $B_p$ is large; e.g., for
NbN films with $d=20$ nm the RF field can be as high as
4.2 T.  From the elastic and viscose forces on a
parallel vortex in a thin film, Gurevich estimates its
characteristic relaxation time as
 \begin{eqnarray}   % 24
 \tau \approx 2d\mu_0 \lambda^2 / (\xi \rho_n) \,.
 \end{eqnarray}
For a 30 nm Nb$_3$Sn film this $\tau \approx 10^{-12}$ s
is much shorter than the RF period of $10^{-9}$ s. The
maximum amplitude of the RF field at which the surface
barrier of a single thin film coating disappears is of
the order of the bulk $H_c$ of the film material, e.g.,
0.54 T for Nb$_3$Sn. Thus, Nb$_3$Sn coating more than
doubles the vortex penetration field for Nb,
$B_p \approx B_c =0.18$ T at low $T$. It appears
that Nb cavities coated with a Nb$_3$Sn layer or with
NbN multilayers allow for much higher RF amplitudes
than uncoated Nb, or Cu coated by a Nb film, if this
can be achieved technically.

\end{document}